\documentstyle[aps,prd]{revtex}
\begin{document}
\hfill{hep-ph/0210173}\par
\hfill{OCHA-PP-193}\par
\vskip 0.5cm
\centerline{\large\bf $k_T$ factorization of exclusive processes}
\vskip 0.3cm
\centerline{Makiko Nagashima$^1$ and Hsiang-nan Li$^2$}
\vskip 0.3cm
\centerline{$^1$Department of Physics, Ochanomizu University,}\par
\centerline{Bunkyo-ku, Tokyo 112-8610, Japan}
\vskip 0.3cm
\centerline{$^2$Institute of Physics, Academia Sinica,
Taipei, Taiwan 115, Republic of China}
\vskip 0.3cm
\centerline{$^2$Department of Physics, National Cheng-Kung University,}\par
\centerline{Tainan, Taiwan 701, Republic of China}
\vskip 1.0cm
\centerline{\bf abstract}
\vskip 0.3cm

We prove $k_T$ factorization theorem in perturbative QCD (PQCD) for 
exclusive processes by considering $\pi\gamma^*\to \gamma(\pi)$ and
$B\to\gamma(\pi) l\bar\nu$. The relevant form factors are expressed as the
convolution of hard amplitudes with two-parton meson wave functions in the
impact parameter $b$ space, $b$ being conjugate to the parton transverse
momenta $k_T$. The point is that on-shell valence partons carry
longitudinal momenta initially, and acquire $k_T$ through collinear gluon
exchanges. The $b$-dependent two-parton wave functions with an 
appropriate path for the Wilson links are gauge-invariant. The hard 
amplitudes, defined as the difference between the parton-level diagrams 
of on-shell external particles and their collinear approximation, are 
also gauge-invariant. We compare the predictions for two-body 
nonleptonic $B$ meson decays derived from $k_T$ factorization (the 
PQCD approach) and from collinear factorization (the QCD factorization
approach). 

\section{introduction}
Both collinear and $k_T$ factorizations are the fundamental tools of
perturbative QCD (PQCD), where $k_T$ denotes parton transverse momenta.
For inclusive processes, consider deeply inelastic scattering
(DIS) of a hadron, carrying a momentum $p$, by a virtual photon, carrying
a momentum $q$. Collinear factorization \cite{Ste} and $k_T$ factorization
\cite{CCH,CE,LRS} apply, when DIS is measured at a large and small Bjorken 
variable $x_B\equiv -q^2/(2p\cdot q)$, respectively. The cross section is 
written as the convolution of a hard subprocess with a hadron distribution 
function in a parton momentum fraction $x$ in the former, and in both $x$ 
and $k_T$ in the latter. When $x_B$ is small, $x \ge x_B$ can reach a small 
value, at which $k_T$ is of the same order of magnitude as the 
longitudinal momentum $x p$, and not negligible. For exclusive processes, 
such as hadron form factors, collinear factorization was developed in 
\cite{BL,ER,CZS,CZ}. The range of a parton momentum fraction $x$, 
contrary to that in the inclusive case, is not experimentally controllable, 
and must be integrated over between 0 and 1. Hence, the end-point region 
with a small $x$ is not avoidable. If there is no end-point singularity 
developed in a hard amplitude, collinear factorization works. If such a 
singularity occurs, indicating the breakdown of collinear factorization, 
$k_T$ factorization should be employed. Since $k_T$ factorization theorem 
was proposed \cite{BS,LS}, there had been wide applications to various 
processes \cite{LFF}. However, a rigorous proof is not yet available.

Based on the concepts of collinear and $k_T$ factorizations, the PQCD 
\cite{LY1,CL,YL,KLS} and QCD factorization (QCDF) \cite{BBNS} 
approaches to exclusive $B$ meson decays have been developed, respectively. 
As applying collinear factorization to the semileptonic decay 
$B\to \pi \ell{\bar \nu}$ at large recoil, an end-point singularity from 
$x\to 0$ was observed \cite{SHB}. Some authors then concluded that PQCD is 
not applicable to these decays even in the heavy quark limit 
\cite{BBNS}. According to the above explanation, this conclusion is 
obviously too strong. We would rather conclude that it is collinear
factorization which fails, and that exclusive $B$ meson 
decays demand $k_T$ factorization. Retaining the dependence on the parton 
transverse momentum $k_T$, and 
resumming the resultant double logarithms $\alpha_s\ln^2 k_T$ into a 
Sudakov form factor \cite{LY1}, the singularity does not exist. PQCD is 
then self-consistent and reliable as an expansion in a small coupling 
constant $\alpha_s$ \cite{TLS,L5,WY}. 

In this paper we shall prove the factorization theorem with the $k_T$ 
dependence included into two-parton meson wave functions and into 
hard amplitudes. In our previous works we have proposed a simple 
all-order proof of collinear factorization theorem for the exclusive 
process $\pi\gamma^*\to \gamma(\pi)$ and $B\to\gamma(\pi) l\bar\nu$ 
up to the two-parton twist-3 level \cite{L1}. The proof of $k_T$ 
factorization theorem follows the similar procedures. We stress that 
it is more convenient to perform $k_T$ factorization in the impact 
parameter $b$ space, in which infrared divergences in radiative 
corrections can be extracted from parton-level diagrams explicitly. 
We shall explain how to construct a gauge-invariant $b$-dependent
meson wave function defined as a nonlocal matrix element with a special 
path for the Wilson link. Evaluating this matrix element in perturbation
theory, the infrared divergences in the parton-level diagrams are 
exactly reproduced. 

We emphasize that predictions for a physical quantity from $k_T$ 
factorization theorem are gauge-invariant, even though three-parton wave 
functions are not included. The valence partons, carrying only longitudinal 
momenta, are initially on-shell. They acquire the transverse degrees of 
freedom through collinear gluon exchanges, before participating hard 
scattering. Therefore, the parton-level amplitudes are gauge-invariant. A 
hard amplitude, derived from the parton-level amplitudes
with the gauge-invariant and infrared-divergent meson wave function 
being subtracted, is then gauge-invariant and infrared-finite. At last, we 
obtain gauge-invariant and infrared-finite predictions by convoluting the 
hard amplitude with a model wave function, which is determined from 
nonperturbative methods (such as lattice QCD and QCD sum rules).

\section{FACTORIZATION OF $\pi\gamma^\ast\to \gamma(\pi)$}

We first prove $k_T$ factorization theorem for the exclusive 
process $\pi\gamma^\ast \to\gamma$. This process, though containing no 
end-point singularity, 
is simple and appropriate for a demonstration. The momentum $P_1\;(P_2)$ 
of the initial-state pion (final-state photon) is chosen as
\begin{eqnarray}
P_1&=&(P_1^+,0,{\bf 0}_T)=\frac{Q}{\sqrt{2}}(1,0,{\bf 0}_T)\;,
\nonumber\\
P_2&=&(0,P_2^-,{\bf 0}_T)=\frac{Q}{\sqrt{2}}(0,1,{\bf 0}_T)\;.
\label{mpp}
\end{eqnarray}
We concentrate on the kinematic region with large $Q=\sqrt{-q^2}$, 
$q=P_2-P_1$ being the momentum transfer from the virtual photon, in 
which scattering mechanism is governed by PQCD. 
The lowest-order diagrams are displayed in Fig.~1. Assume that the 
on-shell $u$ and $\bar u$ quarks carry the fractional momenta 
${\bar x}P_1$ and $xP_1$, respectively, with ${\bar x}\equiv 1-x$. The 
reason for considering an arbitrary $x$ will become clear later. 
Figure~1(a) gives the parton-level amplitude,
\begin{eqnarray}
{\cal G}^{(0)}(x)&=& -ie^2 {\bar u}(xP_1)\not \epsilon 
\frac{\not P_2-x\not P_1}{(P_2-xP_1)^2}\gamma_\mu u({\bar x}P_1),  
\label{ga0} 
\end{eqnarray}
where $\epsilon$ denotes the polarization vector of the outgoing photon.
Figure~1(b) leads to the same result.

The factorization in the fermion flow is achieved by inserting the 
Fierz identity, 
\begin{eqnarray}
I_{ij}I_{lk}&=& \frac{1}{4}I_{ik}I_{lj} 
+ \frac{1}{4}(\gamma^\alpha)_{ik}(\gamma_\alpha)_{lj}
+ \frac{1}{4}(\gamma^5\gamma^\alpha)_{ik}(\gamma_\alpha\gamma^5)_{lj}
+ \frac{1}{4}(\gamma^5)_{ik}(\gamma^5)_{lj}
+ \frac{1}{8}(\gamma^5\sigma^{\alpha\beta})_{ik}
(\sigma_{\alpha\beta}\gamma^5)_{lj},
\label{fierz0}
\end{eqnarray}
with $I$ being the identity matrix and 
$\sigma_{\alpha\beta}\equiv i[\gamma_\alpha,\gamma_\beta]/2$. For the momenta chosen in Eq.~(\ref{mpp}), only the structure 
$\gamma^5\gamma^\alpha$ with $\alpha=+$ contributes to the wave function 
at leading twist (twist 2). The other structures contribute at higher
twists, and the factorization of the corresponding wave functions is 
similar. 

Equation~(\ref{ga0}) is then factorized into
\begin{eqnarray}
{\cal G}^{(0)}(x)=\psi^{(0)}(x){\cal H}^{(0)}(x)\;,
\label{gl0}
\end{eqnarray}
where
\begin{eqnarray}
\psi^{(0)}(x)&=&\frac{1}{4P_1^+}{\bar u}(xP_1)\gamma^5\not n_-
u({\bar x}P_1)\;,
\nonumber\\
{\cal H}^{(0)}(x)&=& ie^2
\frac{tr(\not \epsilon \not P_2 \gamma_\mu
\not P_1\gamma^5)}{2x P_1\cdot P_2}\;,
\label{h0}
\end{eqnarray}
with the dimensionless vector $n_-=(0,1,{\bf 0}_T)$  on the light cone,
define the lowest-order distribution amplitude and  hard amplitude 
in perturbation theory, respectively. Note that none of 
${\cal G}^{(0)}(x)$, $\psi^{(0)}(x)$, and ${\cal H}^{(0)}(x)$
depends on a transverse momentum in the $O(\alpha_s^0)$ factorization. 

\subsection{$O(\alpha_s)$ Factorization}

Next we consider the $O(\alpha_s)$ radiative corrections to Fig.~1(a) 
shown in Figs.~2(a)-2(f), where the gluon carries the loop
momentum $l$. As stated in \cite{L1}, there are two types of infrared 
divergences, soft and collinear, which arise from $l$ with the 
components,
\begin{eqnarray}
& &l^+\sim l^-\sim l_T\sim \bar{\Lambda}\;,
\nonumber\\
& &l^+\sim Q\;,\;\; l^-\sim \bar{\Lambda}^2/Q\;,\;\; 
l_T\sim \bar{\Lambda}\;,
\label{ragl}
\end{eqnarray}
respectively. Here $\bar \Lambda$, being of $O(\Lambda_{\rm QCD})$, 
represents a small scale. Below we work out the factorization of the 
collinear enhancement from $l$ parallel to $P_1$ without integrating out 
the transverse components $l_T$. The prescription is basically similar to
that for collinear factorization. The wave function and 
the hard amplitude then become $l_T$-dependent through collinear gluon 
exchanges. 

We derive the $O(\alpha_s)$ $k_T$ factorization formula,
written as the convolution over the momentum fraction $\xi$ and over 
the impact parameter $b$,
\begin{eqnarray}
{\cal G}^{(1)}(x)&=& \sum_{i=a}^{f}{\cal G}^{(1)}_i(x)\;,
\nonumber\\
{\cal G}^{(1)}_i(x)&=&\int d\xi\frac{d^2b}{(2\pi)^2}
\phi_i^{(1)}(x,\xi,b)H^{(0)}(\xi,b)
+\psi^{(0)}(x){\cal H}_i^{(1)}(x)\;, 
\label{h11}
\end{eqnarray}
The above expression, with the $O(\alpha_s)$ wave functions 
$\phi_i^{(1)}(x,\xi,b)$ and $H^{(0)}(\xi,b)$ specified, defines the 
$O(\alpha_s)$ hard amplitudes ${\cal H}_i^{(1)}(x)$, which do not contain
collinear divergences. It is now obvious why we consider an arbitrary 
$x$ for the parton-level diagrams in Figs.~1 and 2: we can obtain the 
functional form of ${\cal H}_i^{(1)}(x)$ in $x$. Equation (\ref{h11}) is
a consequence of our assertion that partons acquire transverse
degrees of freedom through collinear gluon exchanges:
${\cal H}^{(1)}$, convoluted with the lowest-order $l_T$-independent
$\psi^{(0)}$, is then identical to that in
collinear factorization. As explained later, this consequence is
crucial for constructing gauge-invariant hard amplitudes.

Figures~2(a) and 2(c) are self-energy corrections to the external 
lines. In this case the loop momentum $l$ does not flow through the 
hard amplitude. The $O(\alpha_s)$ wave functions extracted from
these two diagrams are the same as in the collinear 
factorization \cite{L1}. We simply quote the results,
\begin{eqnarray}
\phi^{(1)}_a(x,\xi,b)&=&
\frac{-ig^2C_F}{4P_1^+} \int\frac{d^4l}{(2\pi)^4}{\bar u}(xP_1)
\gamma^5\not n_-\frac{1}{{\bar x}\not P_1}\gamma^\nu
\frac{{\bar x}\not P_1+\not l}{({\bar x}P_1+l)^2}\gamma_\nu
u({\bar x}P_1)\frac{1}{l^2}\delta(\xi-x)\;,
\label{psia}\\
\phi^{(1)}_c(x,\xi,b)&=&
\frac{-ig^2C_F}{4P_1^+} \int\frac{d^4l}{(2\pi)^4}
{\bar u}(xP_1)\gamma^\nu
\frac{x\not P_1-\not l}{(xP_1-l)^2}\gamma_\nu
\frac{1}{x\not P_1}\gamma^5\not n_-
u({\bar x}P_1)\frac{1}{l^2}\delta(\xi-x)\;.
\label{psic}
\end{eqnarray}

The loop integrand associated with Fig.~2(b) is given by
\begin{eqnarray}
I_b^{(1)}&=& e^2g^2C_F {\bar u}(xP_1)\gamma^\nu
\frac{x\not P_1 -\not l}{(xP_1-l)^2}\not \epsilon 
\frac{\not P_2-x\not P_1 +\not l}{(P_2-xP_1+l)^2} \gamma_\mu
\frac{{\bar x}\not P_1+\not l}{({\bar x}P_1+l)^2}\gamma_\nu u({\bar x}P_1) 
\frac{1}{l^2}\;.
\label{i2b}
\end{eqnarray}
Inserting the Fierz identity, we obtain the wave function,
\begin{eqnarray}
\phi^{(1)}_b(x,\xi,b)&=&\frac{ig^2C_F}{4P_1^+}
\int\frac{d^4l}{(2\pi)^4}{\bar u}(xP_1)
\frac{\gamma^\nu(x\not P_1-\not l)\gamma^5\not n_-
({\bar x}\not P_1+\not l)\gamma_\nu} 
{(xP_1-l)^2({\bar x}P_1+l)^2l^2}u({\bar x}P_1)
\delta\left(\xi-x+\frac{l^+}{P_1^+}\right)e^{-i{\bf l}_T\cdot {\bf b}}\;.
\label{p2b}
\end{eqnarray}
The Fourier transformation introduces the additional factor 
$\exp(-i{\bf l}_T\cdot {\bf b})$ into the wave function 
$\phi^{(1)}_b$ compared to the result in collinear factorization
\cite{L1}, since the hard amplitude depends on $l_T$ in this case.

The integrand associated with the two-particle irreducible diagram in 
Fig.~2(d) is given by
\begin{eqnarray}
I_d^{(1)}&=& -e^2g^2C_F {\bar u}(xP_1)\not \epsilon
\frac{\not P_2-x\not P_1}{(P_2-xP_1)^2}\gamma^\nu
\frac{\not P_2-x\not P_1 +\not l}{(P_2-xP_1+l)^2} 
\gamma_\mu
\frac{{\bar x}\not P_1+\not l}{({\bar x}P+l)^2}\gamma_\nu u({\bar x}P_1) 
\frac{1}{l^2}.
\label{i2d}
\end{eqnarray}
To collect the leading contribution, $\gamma^\nu$ and $\gamma_\nu$ 
must be $\gamma^-$ and $\gamma_-=\gamma^+$, respectively. In the collinear 
region the following approximation holds,
\begin{eqnarray}
(\not P_2-x\not P_1)\gamma^\nu(\not P_2-x\not P_1+\not l)
&\approx& 2P_2^\nu\not P_2\;,
\label{ap00}
\end{eqnarray}
where the $l^-$ and $l_T$ terms, being power-suppressed compared to 
$P_2^-$, have been dropped.

The factorization of the collinear enhancement from Figs.~2(d) 
requires the further approximation for the product of the two internal 
quark propagators \cite{L1},
\begin{eqnarray}
\frac{2P_2^\nu}{(P_2-xP_1)^2(P_2-xP_1+l)^2}
\approx \frac{n_-^\nu}{n_-\cdot l}\biggl[\frac{1}{(P_2-xP_1)^2}
-\frac{1}{(P_2-xP_1+l)^2}\biggr]\;,
\label{pi}
\end{eqnarray}
where the numerator $2P_2^\nu$ comes from Eq.~(\ref{ap00}), and the factor
$n_-^\nu/n_-\cdot l$ is exactly the Feynman rule associated with a Wilson
line in collinear factorization. Similarly, we have neglected the 
power-suppressed terms, such as $l^2$ and $xP_1\cdot l$. The first (second)
term on the right-hand side of Eq.~(\ref{pi}) corresponds to the case 
without (with) the loop momentum $l$ flowing through the hard amplitude. 

The above eikonal approximation also applies to Fig.~2(e). Hence, the
extracted $O(\alpha_s)$ wave functions are written as 
\begin{eqnarray}
\phi^{(1)}_{d}(x,\xi,b)&=&
\frac{-ig^2C_F}{4P_1^+}\int\frac{d^4l}{(2\pi)^4}
{\bar u}(xP_1)\gamma^5\not n_-
\frac{ {\bar x}\not P_1+\not l}{({\bar x}P_1+l)^2}
\gamma_\nu u({\bar x}P_1)
\frac{1}{l^2}\frac{n_-^\nu}{n_-\cdot l}
\nonumber\\
& &\times\left[\delta(\xi-x)-\delta\left(\xi-x+\frac{l^+}{P_1^+}\right)
e^{-i{\bf l}_T\cdot {\bf b}}\right]\;,
\label{p2d}\\
\phi^{(1)}_{e}(x,\xi,b) &=&
\frac{ig^2C_F}{4P_1^+}\int\frac{d^4l}{(2\pi)^4}
{\bar u}(xP_1)\gamma_\nu\frac{x\not P_1-\not l}{(xP_1-l)^2}
\gamma^5\not n_-
u({\bar x}P_1)\frac{1}{l^2}\frac{n_-^\nu}{n_-\cdot l}
\nonumber\\
& &\times\left[\delta(\xi-x)-\delta\left(\xi-x+\frac{l^+}{P_1^+}\right)
e^{-i{\bf l}_T\cdot {\bf b}}\right]\;,
\label{p2e}
\end{eqnarray}
where the first (second) term in the brackets is associated with the first
(second) term on the right-hand side of Eq.~(\ref{pi}). Due to the Fourier
transformation, the second terms acquire the additional factor 
$\exp(-i{\bf l}_T\cdot {\bf b})$ compared to the results in
collinear factorization.

Figure~2(f) does not exhibit a collinear enhancement, since the 
radiative gluon gives a self-energy correction to the off-shell 
internal line. Hence, we have $\phi^{(1)}_{f}(x,\xi,b) =0$.
It is easy to observe that the soft divergences cancel among
the $O(\alpha_s)$ radiative corrections. In the soft 
region of $l$ we have $\exp(-i{\bf l}_T\cdot {\bf b})\approx 1$ and 
$l^+\approx 0$, and the two terms in Eqs.~(\ref{p2d}) and (\ref{p2e})
cancel. Similarly, the soft divergences cancel among 
Figs.~2(a)-2(c). This is the reason we discuss only the factorization
of the collinear enhancements. 

The above $O(\alpha_s)$ wave functions can be reproduced by the 
$O(\alpha_s)$ terms of the following nonlocal matrix element in the $b$ 
space,
\begin{eqnarray}
\phi(x,\xi,b)&=&i\int\frac{dy^-}{2\pi }e^{-i\xi P_1^+y^-}
\langle 0|{\bar u}(y)\gamma_5\not n_-
P\exp\left[-ig\int_0^{y}ds\cdot A(s)\right]u(0)
|{\bar u}(xP_1) u({\bar x}P_1)\rangle\;,
\label{pw1}
\end{eqnarray}
with the coordinate $y=(0,y^-,{\bf b})$. The path 
for the Wilson link is composed of three pieces: from 0 to $\infty$ 
along the direction of $n_-$, from $\infty$ to $\infty+{\bf b}$, and from  
$\infty+{\bf b}$ back to $y$ along the direction of $-n_-$ as displayed in 
Fig.~3. We show that the first piece corresponds to the eikonal 
line associated with the first terms in 
Eqs.~(\ref{p2d}) and (\ref{p2e}). Fourier transforming the gauge field
$A(s)$ into ${\tilde A}(l)$, we have
\begin{eqnarray}
-ig\int_0^{\infty}dz
\exp[iz(n_-\cdot l+i\epsilon)]n_-\cdot {\tilde A}(l)
=g\frac{n_-^\alpha}{n_-\cdot l} {\tilde A}_\alpha(l)\;.
\end{eqnarray}
The field ${\tilde A}(l)$, contracted with other gauge fields, gives 
the propagator of the gluon attaching the eikonal line.
The second piece does not contribute because of the appropriate 
choice of the small imaginary constant $+i\epsilon$ in the above
expression. The third piece corresponds to the eikonal line associated
with the second terms in Eqs.~(\ref{p2d}) and 
(\ref{p2e}). The additional Fourier factor $\exp(-i{\bf l}_T\cdot {\bf b})$ 
is a consequence of the shift by ${\bf b}$ from the first piece:
\begin{eqnarray}
-ig\int_{\infty}^{y^-}dz
\exp[iz(n_-\cdot l+i\epsilon)-i{\bf l}_T\cdot {\bf b}]
n_-\cdot {\tilde A}(l)
=-g\frac{n_-^\alpha}{n_-\cdot l}e^{-i{\bf l}_T\cdot {\bf b}}
e^{i l^+y^-}{\tilde A}_\alpha(l)\;,
\end{eqnarray}
where the Fourier factor $\exp(i l^+y^-)$
leads to the function $\delta(\xi-x+l^+/P_1^+)$.

At last, for the evaluation of the lowest-order hard amplitude, we 
neglect only the minus component $l^-$ in the denominator [see the second
term on the right-hand side of Eq.~(\ref{pi})], 
\begin{eqnarray}
(P_2-xP_1+l)^2 \approx -(2\xi P_1\cdot P_2 + l_T^2) \;.
\label{appI}
\end{eqnarray}
Note that in collinear factorization both $l^-$ and $l_T$ are dropped. 
The $b$-dependent hard amplitude is then given by,
\begin{eqnarray}
H^{(0)}(\xi,b)&=&\int d^2l_T
{\cal H}^{(0)}(\xi,l_T)\exp(i{\bf l}_T\cdot {\bf b})\;,
\nonumber\\
{\cal H}^{(0)}(\xi,l_T)&=& ie^2
\frac{tr(\not \epsilon \not P_2\gamma_\mu
\not P_1\gamma^5)}{2\xi P_1\cdot P_2+l_T^2}\;.
\label{psi0}
\end{eqnarray}
Equivalently, the above ${\cal H}^{(0)}(\xi,l_T)$ is derived by
considering an off-shell $\bar u$ quark, which carries the momentum
$\xi P_1-{\bf l}_T$, and the leading structure $\not P_1\gamma_5$
associated with the pion, which is the same as in collinear 
factorization.

\subsection{All-order Factorization}

In this subsection we present the all-order proof of $k_T$ factorization
theorem for the process $\pi\gamma^\ast \to \gamma$, and construct the
parton-level wave function in Eq.~(\ref{pw1}). The proof is similar to 
that for collinear factorization, if it is performed in the impact 
parameter $b$ space. It will be observed that collinear factorization is 
the $b\to 0$ limit of $k_T$ factorization. Therefore, we just highlight 
the differences, and refer the rest of details to \cite{L1}. The idea of 
the proof is based on induction. The factorization of the $O(\alpha_s)$ 
collinear enhancements has been derived in the previous subsection. 
Consider $G^{(0)}(x,b)$ and $G^{(1)}(x,b)$ defined via
\begin{eqnarray}
{\cal G}^{(0),(1)}(x)\equiv{\cal G}^{(0),(1)}(x,k_T=0)=
\int \frac{d^2b}{(2\pi)^2} G^{(0),(1)}(x,b)\;,
\end{eqnarray}
which indicates that the integration over the variable $b$
corresponds to an amplitude with $k_T=0$ for external particles. 
The $O(\alpha_s)$ hard amplitude $H^{(1)}(\xi,b)$ is defined similarly
via ${\cal H}^{(1)}(\xi)$. We obtain the factorization formula up to 
$O(\alpha_s)$,
\begin{eqnarray}
G^{(0)}(x,b) + G^{(1)}(x,b) &=&\int d\xi
\Bigl[\phi^{(0)}(x,\xi,b) + \phi^{(1)}(x,\xi,b) \Bigr]
\Bigl[H^{(0)}(\xi,b) + H^{(1)}(\xi,b) \Bigr]\;,
\label{bfa}
\end{eqnarray}
with $\phi^{(0)}(x,\xi,b)=\psi^{(0)}(x)\delta(\xi-x)$.
The summation over all the diagrams is understood.

Assume that factorization theorem holds up to $O(\alpha_s^N)$,
\begin{eqnarray}
G^{(j)}(x,b)=\sum_{i=0}^{j}\int d\xi 
\phi^{(i)}(x,\xi,b)H^{(j-i)}(\xi,b)\;,\;\;\;\;
j=1,\cdots, N\;,
\label{gbn}
\end{eqnarray} 
where $\phi^{(i)}(x,\xi,b)$ is given by the $O(\alpha_s^{i})$ terms in the
perturbative expansion of Eq.~(\ref{pw1}). $H^{(j-i)}(\xi,b)$ stands for
the $O(\alpha_s^{j-i})$ infrared-finite hard amplitude. 
Equations~(\ref{bfa}) and (\ref{gbn}) approach the expressions in 
collinear factorization as $b\to 0$ as stated above. We shall show that 
the $O(\alpha_s^{N+1})$ diagrams ${\cal G}^{(N+1)}$ in the momentum space 
is written as the convolution of the $O(\alpha_s^N)$ diagrams 
${\cal G}^{(N)}$ with the $O(\alpha_s)$ wave function by employing the 
Ward identity,
\begin{eqnarray}
l_\mu G^\mu(l,k_1,k_2,\cdots, k_n)=0\;,
\label{war}
\end{eqnarray}
where $G^\mu$ represents a physical amplitude with an external gluon
carrying the momentum $l$ and with $n$ external quarks carrying the
momenta $k_1$, $k_2$, $\cdots$, $k_n$. All these external particles are
on the mass shell. It is known that factorization of a QCD process in 
momentum, spin and color spaces requires summation of many 
diagrams. With the Ward identity, the diagram 
summation can be handled in an elegant way.  

Look for the gluon in a complete set of $O(\alpha_s^{N+1})$ diagrams 
${\cal G}^{(N+1)}$, one of whose ends attaches the outer most vertex on 
the upper $u$ quark line in the pion. Let $\alpha$ denote the outer most
vertex, and $\beta$ denote the attachments of the other end of the 
identified gluon inside the rest of the diagrams. There are two 
types of collinear configurations associated with this gluon, depending on 
whether the vertex $\beta$ is located on an internal line with a momentum 
along $P_1$. The quark spinor adjacent to the vertex $\alpha$ is 
$u({\bar x}P_1)$. If $\beta$ is not located on a collinear line along $P_1$, 
the component $\gamma^+$ in $\gamma^\alpha$ and the minus component of 
the vertex $\beta$ give the leading contribution. If $\beta$ is located on 
a collinear line along $P_1$, $\beta$ can not be minus, and both $\alpha$
and $\beta$ label the transverse components. This configuration is the 
same as of the self-energy correction to an on-shell particle.

According to the above classification, we decompose the tensor
$g_{\alpha\beta}$ appearing in the propagator of the identified gluon 
into
\begin{eqnarray}
g_{\alpha\beta}=\frac{n_{-\alpha} l_\beta}{n_-\cdot l}
-\delta_{\alpha \perp}\delta_{\beta \perp}
+\left(g_{\alpha\beta}-\frac{n_{-\alpha} l_\beta}{n_-\cdot l}
+\delta_{\alpha \perp}\delta_{\beta \perp}\right)\;.
\label{dec}
\end{eqnarray}
The first term on the right-hand side extracts the first 
type of collinear enhancements, since the light-like vector $n_{-\alpha}$ 
selects the plus component of $\gamma^\alpha$, and the dominant component
$l_{\beta=-}$ in the collinear region selects the minus component of the 
vertex $\beta$. The components $l_{\beta=+,\perp}$ do not change the 
collinear structure, since they are negligible in the numerators compared 
to the leading terms proportional to $P_1^+$ and $P_2^-$. This can be 
confirmed by contracting $l_\beta$ to Figs.~2(d) and 2(e), from which 
Eq.~(\ref{pi}) is obtained. The second term extracts the second type of 
collinear enhancements. The last term does not contribute a collinear 
enhancement due to the equation of motion for the $u$ quark. We shall 
concentrate on the factorization of ${\cal G}_\parallel^{(N+1)}$ 
corresponding to the first term on the right-hand side of Eq.~(\ref{dec}), 
and the factorization associated with the second term can be included 
simply by following the procedure in  \cite{L1}. 

Those diagrams with Figs.~2(a) and 2(b) as the $O(\alpha_s)$ 
subdiagrams are excluded from the set of ${\cal G}_\parallel^{(N+1)}$ as 
discussing the first type of collinear configurations, since the identified
gluon does not attach a line parallel to $P_1$. Consider the physical 
amplitude, in which the two on-shell quarks and one on-shell gluon carry 
the momenta ${\bar \xi} P_1$, $x P_1$ and $l$, respectively. Figure 4(a), 
describing the Ward identity, contains a complete set of contractions of 
$l_\beta$, since the second and third diagrams have been added back. The
second and third diagrams in Fig.~4(a) lead to
\begin{eqnarray}
& &l_\beta \frac{1}{{\bar \xi}\not P_1-\not l}\gamma^\beta
u({\bar \xi} P_1)
=\frac{1}{{\bar \xi}\not P_1-\not l}(\not l-{\bar \xi}\not P_1 +
{\bar \xi}\not P_1)u({\bar \xi} P_1)
=-u({\bar \xi} P_1)\;,
\label{ide}\\
& &l_\beta{\bar u}(xP_1)\gamma^\beta\frac{1}{x\not P_1-\not l}
=-{\bar u}(xP_1)\;,
\label{ide2}
\end{eqnarray}
respectively. The terms $u({\bar \xi} P_1)$ and ${\bar u}(xP_1)$ at the 
ends of the above expressions correspond to the $O(\alpha_s^N)$ diagrams.

Figure~4(b) shows that the diagrams ${\cal G}_{\parallel}^{(N+1)}$ 
associated with the first term in Eq.~(\ref{dec}) are factorized into the 
convolution of the parton-level $O(\alpha_s^N)$ diagrams ${\cal G}^{(N)}$ 
with the $O(\alpha_s)$ collinear piece extracted from Fig.~2(d). 
The double line represents the Wilson line.
The first diagram means that the gluon momentum does not flow into 
${\cal G}^{(N)}$, while in the second diagram the gluon momentum does. 
The similar reasoning applies to the identified gluon, 
one of whose ends attaches the outer most vertex of the lower $\bar u$ 
quark line. 
Substituting Eq.~(\ref{gbn}) into $G^{(N)}(\xi,b)$ in the $b$ space on
the right-hand side of Fig.~4(b), and 
following the procedure in \cite{L1}, we arrive at
\begin{eqnarray}
G^{(N+1)}(x,b) =\sum_{i=0}^{N+1} \int d\xi 
\phi^{(i)}(x,\xi,b) H^{(N+1-i)}(\xi,b)\;,
\label{gf2}
\end{eqnarray}
with the infrared-finite $O(\alpha_s^{N+1})$ hard amplitude $H^{(N+1)}$.
Equation (\ref{gf2}) implies that all the collinear enhancements in the 
process $\pi\gamma^\ast \to \gamma$ can be factorized into the 
wave function in Eq.~(\ref{pw1}) order by order.

\subsection{Gauge Invariance}

We now demonstrate the gauge invariance of $k_T$ factorization
theorem. Equation~(\ref{pw1}) is explicitly gauge-invariant because 
of the presence of the Wilson link from $0$ to $y$ \cite{CE,CS2}. Below 
we argue that hard amplitudes in $k_T$ factorization are also 
gauge-invariant. Equation (\ref{h11}) approaches the collinear 
factorization under the approximation,
\begin{eqnarray}
\phi^{(1)}(x,\xi,b)\approx \phi^{(1)}(x,\xi,0)\equiv
\psi^{(1)}(x,\xi)\;,
\end{eqnarray}
with $\psi^{(1)}(x,\xi)$ being the distribution amplitude in
collinear factorization. The integration of the hard amplitude 
$H^{(0)}(\xi,b)$ over $b$ gives ${\cal H}^{(0)}(\xi,l_T=0)$. Hence, 
we have the collinear factorization formula,
\begin{eqnarray}
{\cal G}^{(1)}(x)=\int d\xi\psi^{(1)}(x,\xi){\cal H}^{(0)}(\xi)
+\psi^{(0)}(x){\cal H}^{(1)}(x)\;,
\end{eqnarray}
where the summation over the diagrams has been suppressed. 
Since ${\cal G}^{(1)}(x)$, $\psi^{(1)}(x,\xi)$, and
${\cal H}^{(0)}(\xi)$ are gauge-invariant in collinear factorization,
${\cal H}^{(1)}(x)$ is gauge-invariant. From Eq.~(\ref{h11}), the
gauge invariance of $\phi^{(1)}(x,\xi,b)$ stated above, together with the
gauge invariance of ${\cal G}^{(1)}(x)$ and ${\cal H}^{(1)}(x)$, then
imply the gauge invariance of $H^{(0)}(\xi,b)$. Similarly, the $k_T$
factorization formula of $O(\alpha_s^2)$,
\begin{eqnarray}
{\cal G}^{(2)}(x)=\int d\xi\frac{d^2b}{(2\pi)^2}
\left[\phi^{(2)}(x,\xi,b)H^{(0)}(\xi,b)
+\phi^{(1)}(x,\xi,b)H^{(1)}(\xi,b) \right]
+\psi^{(0)}(x){\cal H}^{(2)}(x)\;, 
\end{eqnarray}
leads to the gauge invariance of $H^{(1)}(\xi,b)$:
Both ${\cal G}^{(2)}(x)$ and $\psi^{(0)}(x){\cal H}^{(2)}(x)$
are gauge-invariant in collinear factorization, and all
$\phi^{(i)}(x,\xi,b)$ are gauge-invariant as explained
previously.
The gauge invariance of $H^{(0)}(\xi,b)$ stated above then implies
the gauge invariance of $H^{(1)}(\xi,b)$.
 Therefore, the hard amplitudes in $k_T$ 
factorization are gauge-invariant at all orders.

Equation~(\ref{pw1}) plays the role of an infrared regulator for
parton-level diagrams. A hard amplitude then corresponds to the 
regularized parton-level diagrams. After determining the gauge-invariant
infrared-finite hard amplitude $H(x,b)$, we convolute it with the physical
two-parton pion wave function, whose all-order gauge-invariant definition 
is given by
\begin{eqnarray}
\phi(x,b)&=&i\int\frac{dy^-}{2\pi }e^{-ixP_1^+y^-}
\langle 0|{\bar u}(y)\gamma_5\not n_-
P\exp\left[-ig\int_0^{y}ds\cdot A(s)\right]u(0)|\pi(P_1)\rangle\;.
\label{pwt}
\end{eqnarray}
The valence-quark state $|{\bar u}(xP_1) u({\bar x}P_1)\rangle$
has been replaced by the pion state $|\pi(P_1)\rangle$, and the pion decay 
constant $f_\pi$ has been omitted. The relevant form factor $F$ for the
process $\pi\gamma^\ast \to\gamma$ is then expressed as
\begin{eqnarray}
F=\int dx \frac{d^2b}{(2\pi)^2}\phi(x,b)H(x,b)\;.
\end{eqnarray}
We conclude that predictions derived from $k_T$
factorization theorem are gauge-invariant and infrared-finite.

$k_T$ factorization theorem for the pion form factor involved in the
process $\pi\gamma^*\to \pi$ can be proved in the same way.
The $O(\alpha_s)$ factorization is similar to the collinear factorization
performed in \cite{L1}. The only difference is the extra Fourier factor
$\exp(-i{\bf l}_T\cdot {\bf b})$ associated with the diagrams, in which 
the loop momentum flows through the hard amplitude. Following
the steps in Sec.~II A, the eikonal line can be constructed from the 
diagrams with collinear gluons attaching the hard amplitude and the 
outgoing pion. The decomposition in Eq.~(\ref{dec}) and the whole 
procedure presented above then apply. That is, the all-order proof is also
similar to that of collinear factorization \cite{L1}. Compared to the
process $\pi\gamma^*\to \gamma$, the structures $\gamma_5$ and 
$\gamma_5\sigma^{\alpha\beta}$ from the Fierz identity 
contribute, and the corresponding twist-3 pion wave functions
appear.

\section{FACTORIZATION of $B\to\gamma(\pi)\ell\bar\nu$}

In this section we prove $k_T$ factorization theorem for the radiative 
decay $B\to \gamma l{\bar \nu}$, retaining the transverse degrees of 
freedom of internal particles, and construct the $B$ meson wave function 
in the impact parameter $b$ space. We shall discuss only the $O(\alpha_s)$
factorization, and demonstrate that the all-order factorization can be
proved in a way similar to collinear factorization \cite{L1}.
The momentum $P_1$ of the $B$ meson and the momentum $P_2$ of the 
out-going on-shell photon are chosen as
\begin{eqnarray}
P_1 &=& \frac{M_B}{\sqrt 2}\;(1,1,{\bf 0}_T)\;, \;\;\;
P_2 = \frac{M_B}{\sqrt 2}\;(0,\eta,{\bf 0}_T)\;, 
\end{eqnarray}
where the photon energy fraction $\eta$ is large enough to justify the
applicability of PQCD. Assume that the light spectator quark in the $B$
meson carries the momentum $k$. In collinear factorization, only the plus
component $k^+$ is relevant through the inner product $k\cdot P_2$
\cite{DS}. The lowest-order diagrams for the $B\to \gamma\ell\bar\nu$
decay is displayed in Fig.~1, but with the upper quark (virtual photon) 
replaced by a $b$ quark ($W$ boson).

Bellow we shall concentrate on Fig.~1(a), because Fig.~1(b) is
power-suppressed. Figure~1(a) gives the parton-level amplitude,
\begin{eqnarray}
{\cal G}^{(0)}(x)&=&
e{\bar u}(k)\not \epsilon \frac{\not P_2-\not k}{(P_2-k)^2}
\gamma_\mu(1-\gamma_5) b(P_1-k)\;,
\label{b1a}
\end{eqnarray}
which does not depend on a transverse momentum.
Inserting the Fierz identity in Eq.~(\ref{fierz0}) into to Eq.~(\ref{b1a}),
we obtain Eq.~(\ref{gl0}) with
\begin{eqnarray}
\psi^{(0)}(x)&=&\frac{1}{4P_1^+}{\bar u}(k)\gamma_5\not n_-
b(P_1-k)\;,
\nonumber\\
{\cal H}^{(0)}(x)&=&-e
\frac{tr[\not \epsilon \not P_2 \gamma_\mu(1-\gamma_5)
\not n_+\gamma^5]P_1^+}{2x P_1\cdot P_2}\;,
\nonumber\\
&=&-e\frac{tr[\not \epsilon \not P_2 \gamma_\mu(1-\gamma_5)
(\not P_1+M_B)(\not n_+/\sqrt{2})\gamma^5]}{2x P_1\cdot P_2}\;,
\label{h0B}
\end{eqnarray}
with the dimensionless vector $n_+=(1,0,{\bf 0}_T)$ on the light cone.
We have dropped the higher-power term $\not k$ in the numerator,
and the momentum fraction $x$ is defined by $x=k^+/P_1^+$. For the $B$
meson wave functions, there are two leading-twist components associated 
with the structures $\gamma_5\gamma^\pm$. For the $B\to\gamma l\bar\nu$ 
decay, we choose the structure $\gamma_5\gamma^+=\gamma_5\not n_-$,
since $\not \epsilon$ in Eq.~(\ref{h0B}) involves $\gamma_\perp$, 
and only the structure $\gamma^-\gamma_5=\not n_+\gamma_5$ contributes to
the hard amplitude.

Next we consider the $O(\alpha_s)$ radiative corrections to Fig.~1(a)
shown in Figs.~2(a)-2(f). We discuss the factorization
of the soft divergence from the loop momentum
$l^\mu\sim ({\bar \Lambda},{\bar \Lambda}, {\bar \Lambda})$,
where $\bar\Lambda$ can be regarded as the $B$ meson and $b$ quark
mass difference, $\bar\Lambda=M_B-m_b$. The dependence of the $B$ meson 
wave function on the transverse momentum is generated by 
soft gluon exchanges. The analysis is similar to that in Sec.~II, and 
we obtain Eq.~(\ref{h11}). The factorization of the two-particle reducible
diagrams in Fig.~2(a)-(c) is straightforward. Take Fig.~2(b) as an 
example, which gives the integrand,
\begin{eqnarray}
I_b^{(1)}&=& ieg^2C_F {\bar u}(k)\gamma^\nu
\frac{\not k -\not l}{(k-l)^2}\not \epsilon 
\frac{\not P_2-\not k +\not l}{(P_2-k+l)^2} \gamma_\mu(1-\gamma_5)
\frac{\not P_1-\not k +\not l+m_b}{(P_1-k+l)^2-m_b^2}
\gamma_\nu b(P_1-k) 
\frac{1}{l^2}\;.
\label{i2bB}
\end{eqnarray}
Employing the eikonal approximation in the heavy-quark limit, we have
\begin{eqnarray}
\frac{\not P_1-\not k+\not l+m_b}{(P_1-k+l)^2-m_b^2}
\gamma_\nu b(P_1-k) &\approx& 
\frac{v_\nu}{v\cdot l}b(P_1-k)\;,
\label{appII}
\end{eqnarray}
with the velocity $v=P_1/M_B$. The $O(\alpha_s)$ wave function 
extracted from Eq.~(\ref{i2bB}) is then written as
\begin{eqnarray}
\phi^{(1)}_{b}(x,\xi,b)&=&\frac{ig^2C_F}{4P_1^+}
\int\frac{d^4l}{(2\pi)^4}{\bar u}(k)
\frac{\gamma^\nu(\not k-\not l)}{(k-l)^2 l^2}
\gamma_5\not n_-b(P_1-k)\frac{v_\nu}{v\cdot l}
\delta\left(\xi-x+\frac{l^+}{P_1^+}\right)
e^{-i{\bf l}_T\cdot {\bf b}}\;.
\label{p2bB}
\end{eqnarray}

The loop integrands associated with Figs.~2(d) and 2(e) 
are given by
\begin{eqnarray}
I_d^{(1)}&=& -ieg^2C_F {\bar u}(k)\not \epsilon 
\frac{\not P_2 -\not k}{(P_2-k)^2}\gamma^\nu
\frac{\not P_2-\not k +\not l}{(P_2-k+l)^2} \gamma_\mu(1-\gamma_5)
\frac{\not P_1-\not k +\not l+m_b}{(P_1-k+l)^2-m_b^2}
\gamma_\nu b(P_1-k) 
\frac{1}{l^2}\;,
\label{i2dB}
\\
I_e^{(1)}&=& ieg^2C_F {\bar u}(k)\gamma_\nu
\frac{\not k -\not l}{(k-l)^2}
\not \epsilon 
\frac{\not P_2 -\not k+\not l}{(P_2-k+l)^2}\gamma^\nu
\frac{\not P_2-\not k}{(P_2-k)^2} 
\gamma_\mu(1-\gamma_5)b(P_1-k) 
\frac{1}{l^2}\;,
\label{i2eB}
\end{eqnarray}
respectively. Neglecting the subleading terms proportional to $\not k$
and $\not l$ in the numerators in comparison with $\not P_2$,
we have the eikonal approximation,
\begin{eqnarray}
\frac{\not P_2-\not k}{(P_2-k)^2}\gamma^\nu
\frac{\not P_2-\not k+\not l}{(P_2-k+l)^2}
\approx
\frac{n_-^\nu }{n_-\cdot l}
\Biggl[ \frac{1}{(P_2-k)^2}- \frac{1}{(P_2-k+l)^2} \Biggr]\not P_2\;,
\end{eqnarray}
similar to Eq.~(\ref{pi}).
Inserting the Fierz identity, we extract the $O(\alpha_s)$ wave functions,
\begin{eqnarray}
\phi^{(1)}_{d}(x,\xi,b)&=&
\frac{-ig^2C_F}{4P_1^+}\int\frac{d^4l}{(2\pi)^4}
{\bar u}(xP_1)\gamma_5\not n_- b(P_1-k)
\frac{1}{l^2}\frac{n_-\cdot v}{n_-\cdot l v\cdot l}
\nonumber\\
& &\times\left[\delta(\xi-x)-\delta\left(\xi-x+\frac{l^+}{P_1^+}\right)
e^{-i{\bf l}_T\cdot {\bf b}}\right]\;,
\label{p2dB}\\
\phi^{(1)}_{e}(x,\xi,b) &=&
\frac{ig^2C_F}{4P_1^+}\int\frac{d^4l}{(2\pi)^4}
{\bar u}(xP_1)\gamma_\nu\frac{\not k-\not l}{(k-l)^2}
\gamma_5\not n_- b(P_1-k)
\frac{1}{l^2}\frac{n_-^\nu}{n_-\cdot l}
\nonumber\\
& &\times\left[\delta(\xi-x)-\delta\left(\xi-x+\frac{l^+}{P_1^+}\right)
e^{-i{\bf l}_T\cdot {\bf b}}\right]\;.
\label{p2eB}
\end{eqnarray}
The eikonal approximation in Eq.~(\ref{appII}) has been
applied. Figure~2(f) does not have the soft divergence due to the
off-shell internal quark.
 
It is obvious that the above $O(\alpha_s)$ parton-level wave functions
are similar to those derived in Sec.~II: the eikonal line in $n_-$ is 
the same as in collinear factorization, and the integrands contain the additional Fourier factor $\exp(-i{\bf l}_T\cdot {\bf b})$, when the
loop momentum flows through the hard amplitude. The decomposition in 
Eq.~(\ref{dec}) and the procedure for the all-order proof presented in 
Sec.~II apply to the $B\to \gamma l{\bar \nu}$ decay. We construct a 
gauge-invariant light-cone $B$ meson wave function,
\begin{eqnarray}
\phi_+(x,b)&=&i\int\frac{dy^-}{2\pi }e^{-ix P_1^+y^-}
\langle 0|{\bar u}(y)\gamma_5\gamma^+
P\exp\left[-ig\int_0^{y}ds\cdot A(s)\right]b_v(0)
|B(P_1)\rangle\;,
\end{eqnarray}
where $b_v$ is the rescaled $b$ quark field characterized by the velocity
$v$. The lowest-order hard amplitude in the $b$ space is given by
Eq.~(\ref{psi0}) with
\begin{eqnarray}
{\cal H}^{(0)}(\xi,l_T)&=&-e
\frac{tr[\not \epsilon \not P_2 \gamma_\mu(1-\gamma_5)
(\not P_1+M_B)(\not n_+/\sqrt{2})\gamma^5]}{2\xi P_1\cdot P_2+l_T^2}\;,
\end{eqnarray}
where the momentum fraction $\xi$ is defined by $\xi= (k^+-l^+)/P_1^+$.
The above expression can be derived by
considering an off-shell $\bar u$ quark of the momentum
$(\xi P_1^+,0,-{\bf l}_T)$, and the leading structure 
$(\not P_1+M_B)(\not n_+/\sqrt{2})\gamma^5$ associated with the
$B$ meson, which is the same as in collinear factorization.

As emphasized in the Introduction, the semileptonic decay
$B\to\pi l\bar\nu$, because of the end-point singularities (the failure
of collinear factorization), demands $k_T$ factorization. Its all-order
proof is also performed in the same way. Note that for this mode, 
both the leading-twist $B$ meson wave functions $\phi_\pm$, associated 
with the structures $\gamma_5\gamma^\pm$, contribute \cite{L1}.

\section{DISCUSSION}

We have explained that the range of a parton momentum
fraction $x$ in exclusive processes, contrary to that in inclusive 
processes, is not experimentally controllable. Hence, the end-point 
region with a small $x$ is not avoidable. If a hard amplitude develops 
an end-point singularity in collinear factorization, implying the
importance of the end-point region, $k_T$ factorization must be employed.
Exclusive $B$ meson decays belong to this category, for which
$k_T$ factorization is a more appropriate tool. We have proved $k_T$
factorization theorem for the processes $\pi\gamma^*\to \gamma(\pi)$ and
$B\to\gamma(\pi) l\bar\nu$ in this paper. The proof performed in the 
impact parameter $b$ space indicates that collinear factorization is the 
$b\to 0$ limit of $k_T$ factorization. 

The prescriptions for determining wave functions and
hard amplitudes in $k_T$ factorization theorem are summarized as 
follows:

$\bullet$ A two-parton $b$-dependent wave function is factorized 
from parton-level diagrams in a way the same as in collinear 
factorization (for example, under the same eikonal approximation), but the loop integrand is associated with an additional Fourier factor 
$\exp(-i{\bf l}_T\cdot {\bf b})$, when the loop momentum $l$ flows through 
a hard amplitude.

$\bullet$ A $k_T$-dependent hard amplitude is obtained in a way the same
as in collinear factorization, but considering off-shell external
partons, which carry the fractional momenta $k=xP+{\bf k}_T$ ($k^2=-k_T^2$),
$P$ being the external meson momenta. Then Fourier transform this hard 
amplitude into the $b$ space.

$\bullet$ The insertion of the Fierz identity to separate the fermion
flow between a wave function and a hard amplitude is the same as
in collinear factorization. Take the process $\pi\gamma^*\to\pi$
discussed in Sec.~II as an example. Up to the twist-3 accuracy for
the initial pion, adopt the structures $\gamma_5\gamma^+$, $\gamma_5$ 
and $\gamma_5\sigma^{\alpha\beta}$ with $\alpha,\beta=\pm$, {\it i.e.}, 
without the $\perp$ components.

Under the above prescriptions, the Wilson link for the $b$-dependent wave
function is the same as in collinear factorization, but with a shift 
${\bf b}$ between the two pieces of paths along the light cone. Both the
$b$-dependent two-parton meson wave functions and hard amplitudes are 
gauge-invariant in $k_T$ factorization, without introducing three-parton 
wave functions. Therefore, predictions for a physical quantity obtained 
from $k_T$ factorization theorem are gauge-invariant. For inclusive
processes in small $x_B$ physics, the gauge invariance of the 
unintegrated gluon distribution function and of the hard subprocess 
of reggeized gluons, being also off-shell by $-k_T^2$, is ensured in a 
similar way. The distinction is that the structures of $\gamma$-matrices 
from the Fierz identity are replaced by eikonal vertices, which contain 
only the longitudinal components \cite{CE}. 
 
There are more differences between the $k_T$ factorizations of inclusive
and exclusive processes. Inclusive processes involve a single scale, 
and only single logarithms. Exclusive processes involve two scales (when 
a valence parton is soft, another is fast), and double logarithms. That 
is, no rapidity ordering is assumed \cite{LL}. Hence, the required 
resummation techniques are different. The definition of meson wave 
functions constructed in this work serves as the starting point of $k_T$
resummation \cite{CS,L96,MR}. The resultant Sudakov factor smears the
end-point singularity in the semileptonic decay $B\to\pi l\bar\nu$ by 
increasing the magnitude of $k_T$ though infinite many gluon exchanges.
The perturbative expansion of decay amplitudes then makes sense. 
Certainly, this conclusion needs to be justified by evaluating 
next-to-leading-order corrections in $\alpha_s$ in $k_T$ factorization
theorem. If higher-order contributions converge quickly enough, the 
PQCD approach to exclusive $B$ meson decays will be theoretically solid.

In our next work we shall construct $k_T$ factorization of
two-body nonleptonic $B$ meson decays. Below we briefly compare the  
phenomenological consequences for these
decays derived from collinear and $k_T$ factorizations,
mentioning only the CP asymmetry in
$B_d^0\to\pi^+\pi^-$ mode. According to the power counting rules of
QCDF \cite{BBNS} based on collinear factorization, the
factorizable emission diagram in Fig.~5(a) gives the leading contribution
of $O(\alpha_s^0)$, since the $B\to\pi$ form factor $F^{B\pi}$ is not
calculable. Because Fig.~5(a) is real, the strong phase arises from the
factorizable annihilation diagram in Fig.~5(b), being of
$O(\alpha_s m_0/M_B)$, and from the vertex correction in Fig.~5(c),
being of $O(\alpha_s)$. For $m_0/M_B$ slightly smaller than unity,
Fig.~5(c) is the leading source of strong phases in collinear
factorization (QCDF). In $k_T$ factorization the power counting rules
change. The factorizable emission diagram is calculable and of
$O(\alpha_s)$ as indicated in Fig.~5(d). The factorizable annihilation
diagram has the same power counting as for Fig.~5(b). The vertex
correction becomes of $O(\alpha_s^2)$ as shown in Fig.~5(e).
Therefore, Fig.~5(b) contributes the leading strong phase in $k_T$
factorization (PQCD). The strong phases from Fig.~5(b) and 5(c)
are opposite in sign, and the former has a large magnitude. This is the
reason QCDF prefers a small and positive CP asymmetry $C_{\pi\pi}$
\cite{Ben}, while PQCD prefers a large and negative
$C_{\pi\pi}\sim -30\%$ \cite{KLS,LUY,Keum}. It is expected that
in the near future the two different approaches to exclusive $B$ meson 
decays, based on collinear and $k_T$ factorizations, could be 
distinguished by experiments \cite{Nir,Ros}.

\vskip 0.3cm
We thank J. Kodaira, Y. Koike, T. Morozumi, G. Sterman, and K. Tanaka for 
useful discussions. The work was supported in part by the National Science
Council of R.O.C. under Grant No. NSC-91-2112-M-001-053, by the National
Center for Theoretical Sciences of R.O.C., and by Theory Group of KEK,
Japan.

\noindent
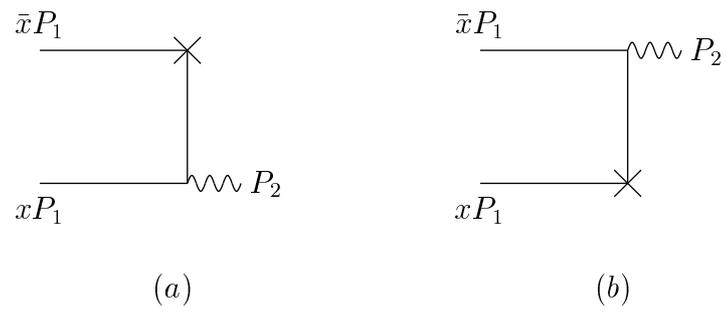
{\bf FIG. 1} Lowest-order diagrams for the process 
$\pi\gamma^\ast \to\gamma$.

\noindent
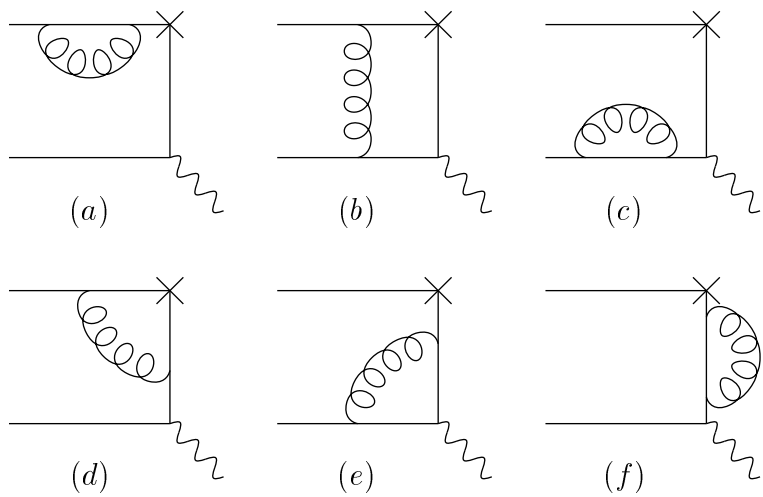
{\bf FIG. 2} $O(\alpha_s)$ corrections to Fig.~1(a).

\noindent
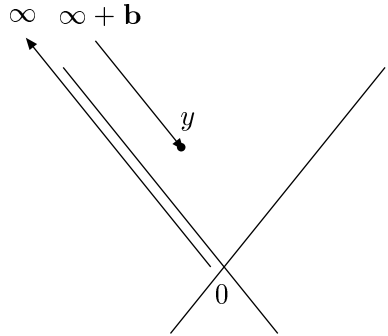
{\bf FIG. 3} The path for the Wilson link in a $b$-dependent
two-parton meson wave function.

\noindent
{\bf FIG. 4} (a) Ward identity. (b) Factorization of ${\cal G}^{(N+1)}$.

\noindent
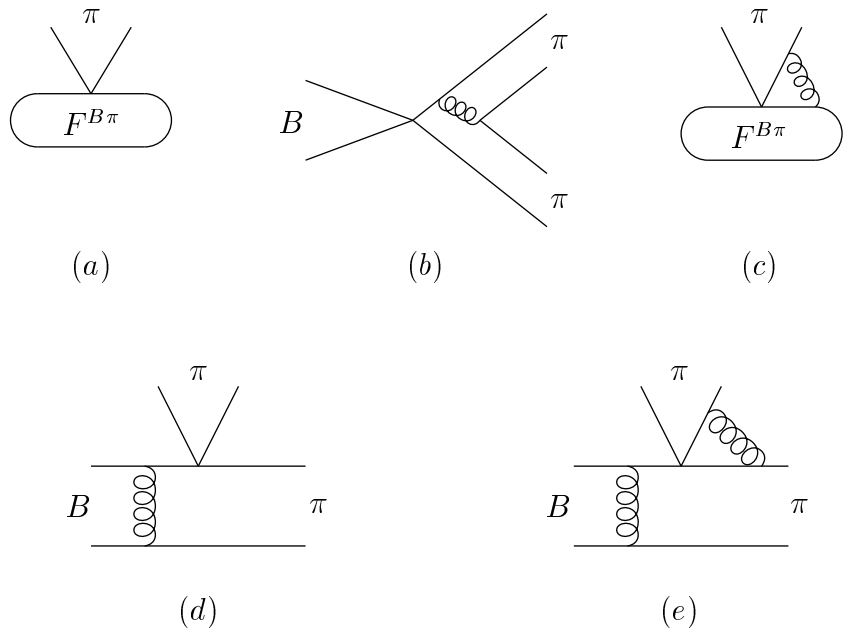
{\bf FIG. 5} Diagrams contributing to the $B_d^0\to\pi^+\pi^-$ decay.

\end{document}